\documentclass{jfm}

\usepackage{graphicx}
\usepackage{natbib}
\usepackage{upmath}
\usepackage{amsmath}
\usepackage{amssymb}
\usepackage{amsbsy}
\usepackage{latexsym}
\usepackage{bm}
\usepackage[utf8]{inputenc}
\usepackage[T1]{fontenc}
\usepackage{upgreek}
\usepackage{xcolor}
\usepackage[caption=false]{subfig}
\usepackage[pdftex]{hyperref}
\usepackage{xspace}

\newcommand{\Canum}{\mbox{\textit{Ca}}\xspace}
\newcommand{\Renumin}{\mbox{\textit{Re}}_0\xspace}
\newcommand{\Renum}{\mbox{\textit{Re}}\xspace}
\newcommand{\Renump}{\ensuremath{\mbox{\textit{Re}}_\mathrm{p}}\xspace}
\newcommand{\visc}{\ensuremath{\eta_\text{a}}}
\newcommand{\viscloc}{\ensuremath{\eta_\text{loc}}}
\newcommand{\depletion}{\ensuremath{d}}
\newcommand{\modb}{\ensuremath{\kappa_\text{b}}}
\newcommand{\mods}{\ensuremath{\kappa_\text{s}}}
\newcommand{\moda}{\ensuremath{\kappa_\alpha}}

\begin{document}

\title[Interplay of inertia and deformability]{Interplay of inertia and deformability on rheological properties of a
suspension of capsules}

\author[Timm~Krueger, Badr~Kaoui and Jens~Harting]%
{Timm~Krueger$^{1,2}$%
  \thanks{Email address for correspondence: timm.krueger@ed.ac.uk},\ns
Badr~Kaoui$^{3,4}$\break and Jens~Harting$^{4,5}$}

\affiliation{$^1$School of Engineering, University of Edinburgh, The King's Buildings, Mayfield Road, EH9 3JL Edinburgh, United Kingdom\\[\affilskip]
$^2$Centre for Computational Science, University College London, 20 Gordon Street, WC1H 0AJ London, United Kingdom\\[\affilskip]
$^3$Physikalisches Institut, Universität Bayreuth, Theoretische Physik I, 95440 Bayreuth, Germany\\[\affilskip]
$^4$Department of Applied Physics, Eindhoven University of Technology, P. O. Box 513, 5600 MB Eindhoven, The Netherlands\\[\affilskip]
$^5$Faculty of Science and Technology, Mesa+ Institute, University of Twente, 7500 AE Enschede, The Netherlands
}

\pubyear{????}
\volume{???}
\pagerange{???--???}
\date{?; revised ?; accepted ?. - To be entered by editorial office}

\maketitle

\begin{abstract}
The interplay of inertia and deformability has a substantial impact on the
transport of soft particles suspended in a fluid.
However, to date a thorough understanding of these systems is still missing and only a limited number of experimental and theoretical studies is available.
We combine the finite-element, immersed-boundary and lattice-Boltzmann methods to simulate three-dimensional suspensions of soft particles subjected to planar Poiseuille flow at finite Reynolds numbers.
Our findings confirm that the particle deformation and inclination increase when inertia is present.
We observe that the Segré-Silberberg effect is suppressed with respect to the
particle deformability.
Depending on the deformability and strength of inertial effects, inward or outward lateral migration of the particles takes place.
In particular, for increasing Reynolds numbers and strongly deformable particles, a hitherto unreported distinct flow focusing effect emerges which is accompanied by a non-monotonic behaviour of the apparent suspension viscosity and thickness of the particle-free layer close to the channel walls.
This effect can be explained by the behaviour of a single particle and the change of the particle collision mechanism when both deformability and inertia effects are relevant.
\end{abstract}


\section{Introduction}
\label{sec_intro}

Fluid inertia and particle deformability --- quantified by the Reynolds (\Renum) and capillary (\Canum) numbers, respectively --- have significant impact on the
individual and collective motion of particles in suspension.
For example, most deformable particles in a viscous flow tend to migrate laterally \emph{towards} the centreline of a Poiseuille flow, creating a depletion layer near the confining walls. However, it is also known that deformable droplets \citep{leal_particle_1980} and vesicles \citep{alexander_farutin_analytical_2013} can migrate \emph{away} from the centreline under certain conditions. Rigid particles at intermediate Reynolds numbers which are originally located near the centreline migrate towards the walls until the inertial lift is balanced by the wall repulsion.
While most studies in the literature focus on the effect of either inertia or deformability separately, we address their combination and as such aim at answering the following question: \emph{what is the interplay between inertia and deformability in a flowing suspension?}  

A deep understanding of the flow of suspensions and their components is of
fundamental practical importance.
For example, lateral migration of particles is exploited for separating and sorting  out cancer cells from healthy blood cells in lab-on-chip devices \Citep{hur_deformability-based_2011, tanaka_separation_2012, geislinger_separation_2012, kruger_numerical_2013}.
The macroscopic viscosity of a suspension, which is of relevance for systems like waste water sewage and industrial production lines, depends on its microscopic composition and particle distribution.
The ultimate goal is to predict the rheology of a suspension based on its microscopic properties, such as local particle concentration and particle
deformation.

About 50 years ago, \citet{segre_behaviour_1962, segre_behaviour_1962-1} discovered that rigid particles in inertial tube flow (with tube radius $r$) have equilibrium positions at a radial distance of about $0.6\, r$.
This equilibrium lateral position results from balancing the confinement
induced repulsive wall force and shear rate gradients.
\Citet{matas_inertial_2004} later extended the experiments to larger Reynolds numbers and observed the formation of an additional inner annulus.
Particles initially located in off-equilibrium positions tend to show cross-streamline migration until they reach an equilibrium position \citep{humphry_axial_2010}.
Understanding the lateral migration of rigid particles in the presence of
inertia is a difficult problem.
The existing asymptotic theory \citep{schonberg_inertial_1989, asmolov_inertial_1999} explaining this behaviour is only valid for weak confinement and small particle Reynolds numbers, although the channel Reynolds number can be large.
 
Numerical lattice-Boltzmann simulations of a single particle and dilute suspensions in the regime $100 \leq \Renum \leq 1000$ showed that multi-body interactions play a crucial role in the formation of inner equilibrium positions \citep{chun_inertial_2006}.
In strongly confined rectangular flows, rigid particles experience inertial focusing and are aligned in one or multiple lines, depending on the particle volume fraction and the channel aspect ratio \citep{humphry_axial_2010}.
These findings demonstrate that there are different mechanisms behind the structuring of particles in the flow which are not always well understood.
The problem even increases in its complexity when considering soft particles.
Flows in microfluidic devices are commonly believed to be in the fully viscous regime (Stokes limit).
However, inertia can play a role if channel diameters are larger than $\approx 100\, \upmu\text{m}$ \citep{di_carlo_inertial_2009}.
Inertia effects may then be used to enhance mixing and to separate or focus
particles in straight or curved geometries \citep{munn_blood_2008, hur_deformability-based_2011, tanaka_separation_2012, martel_inertial_2012}.

The non-inertial lateral motion of single deformable particles (such as vesicles, capsules or red blood cells) in Poiseuille flow has been thoroughly studied over the past years \citep{doddi_lateral_2008, kaoui_lateral_2008, kaoui_lateral_2009, danker_vesicles_2009, coupier_noninertial_2008, shi_lateral_2012}.
Only recently, the effect of inertia on the dynamics of deformable objects has been investigated systematically \citep{doddi_effect_2008, nourbakhsh_three-dimensional_2011, kilimnik_inertial_2011, shin_inertial_2011, kim_numerical_2012, shin_dynamics_2012, laadhari_vesicle_2012, salac_reynolds_2012, luo_inertia-dependent_2013, chen_inertia-_2014}.
Almost all authors report an increase of particle deformation and inclination with \Renum.
Additionally, \citet{doddi_effect_2008} observed a change of the nature of hydrodynamic interactions of capsules upon an increase of the Reynolds number; the self-diffusive collision for small \Renum is replaced by spiralling motion.
Furthermore, \citet{laadhari_vesicle_2012} and \citet{kim_numerical_2012} found that increasing fluid inertia shifts the tank-treading-to-tumbling transition of vesicles to larger values of the viscosity contrast.
\Citet{shin_inertial_2011, shin_dynamics_2012} demonstrated that the lateral equilibrium position of single capsules in 2D inertial channel flow has a maximum at about $\Renum = 50$.

In the present study, we investigate --- \emph{via} combined finite element, immersed boundary and lattice-Boltzmann simulations --- the behaviour of a semi-dilute suspension ($\approx 10\%$ volume fraction) of deformable capsules in a confined channel.
In contrast to previous studies, we consider a large range of channel Reynolds numbers between 3 and 417 and capillary numbers between $0.003$ (nearly rigid) and $0.3$ (strongly deformed).
We investigate in detail the macroscopic and microscopic behaviour of the suspension as function of \Renum and \Canum, such as global and local viscosity, concentration profiles, depletion layer thickness, particle deformation and inclination.
We observe a strong inertial focusing at large \Renum and \Canum; particles tend to cluster in a narrow region about the centreplane, reducing the overall viscosity of the suspension.
This effect is rationalised in terms of the behaviour of a single particle (dilute suspension limit) and hydrodynamic interactions between particles at finite volume fraction.

The article is structured as follows.
Section \ref{sec_methods} briefly describes the numerical model and introduces the simulation parameters.
The numerical results are presented and discussed in section \ref{sec_results_discussion}.
Finally, the findings are summarised and conclusions are provided in section
\ref{sec_summary}.

\section{Computational method and simulation parameters}
\label{sec_methods}

\subsection{Computational model}

We use the Bhatnagar-Gross-Krook (BGK) lattice-Boltzmann (LB) method with the D3Q19 velocity set \citep{succi_lattice_2001, aidun_lattice-boltzmann_2010, kruger_efficient_2011} to compute the fluid flow in the entire domain.
The fluids inside and outside the capsules are modelled with the same density and viscosity.
The kinematic viscosity $\nu$ is related to the BGK relaxation parameter $\tau$ according to
\begin{equation}
 \nu = c_\text{s}^2 \left(\tau - \tfrac{1}{2}\right) \Delta t,
\end{equation}
where $\Delta t$ is the time step and $c_\text{s}$ is the speed of sound.
The lattice constant is denoted $\Delta x$.
The no-slip boundaries at the channel walls are realised through the half-way bounce-back boundary condition \citep{ladd_numerical_1994}.
To include a body force we follow the Shan-Chen forcing approach \citep{shan_multicomponent_1995}.

Each capsule is described as a closed massless membrane discretised into a triangular mesh with 980 facets \citep{kruger_efficient_2011}.
In the absence of external stresses the particles assume a spherical equilibrium shape with radius $r$.
The in-plane energy of a particle membrane is given by the elastic law of \citet{skalak_strain_1973}:
\begin{equation}
 E_\text{s} = \oint \text{d}A\, \left[\frac{\mods}{12} (I_1^2 + 2I_1 -2I_2) + \frac{\moda}{12}I_2^2 \right],
\end{equation}
where $\mods$ is the shear elasticity and $\moda$ is the area dilation modulus and the integral runs over the entire capsule surface.
The parameters $I_1$ and $I_2$ are the in-plane strain invariants which can be derived from the local membrane deformation tensor as detailed by \citet{kruger_efficient_2011}.
To avoid buckling \citep{kilimnik_inertial_2011}, the capsules also have a finite bending resistance with discretised bending energy
\begin{equation}
 E_\text{b} = \frac{\sqrt 3 \modb}{2} \sum_{\langle i, j \rangle} \left(\theta_{ij} - \theta^{\text{eq}}_{ij}\right)^2.
\end{equation}
This is a simplified version of the classical Helfrich form \citep{helfrich_elastic_1973}.
The bending modulus is denoted by $\modb$, and the sum runs over all pairs of neighbouring facets (\emph{i.e.}, facets with one common edge).
Each facet pair has an equilibrium normal-to-normal angle $\theta^{\text{eq}}_{ij}$ defined by the initially spherical capsule shape.
The elastic forces are computed \emph{via} the principle of virtual work \citep{charrier_free_1989, kruger_efficient_2011}.

For the sake of simplicity and reduction of the parameter space dimensionality, the particle elasticities are fixed in a way such that the reduced area dilation modulus and reduced bending modulus,
\begin{equation}
 \tilde \kappa_\alpha = \frac{\kappa_\alpha}{\mods} = 2 \quad \text{and} \quad \tilde \kappa_\text{b} = \frac{\modb}{\mods r^2} = 2.87 \cdot 10^{-3},
\end{equation}
are constant. The typical maximum area extensions following from our choice of $\tilde \kappa_\alpha$ are $<1\%$ for $\Canum = 0.003$, $<8\%$ for $\Canum = 0.03$ and $<20\%$ for $\Canum = 0.3$. The above value for the bending resistance $\modb$ has been chosen for convenience. On the one hand, it is sufficiently small so that the elastic behaviour is dominated by the shear resistance. On the other hand, it is large enough to avoid buckling.

The immersed-boundary method \citep{peskin_immersed_2002} with a trilinear interpolation stencil as presented by \citet{kruger_efficient_2011} is used to couple the fluid flow and capsule dynamics.
The system is assumed to be athermal, and thermal fluctuations are neglected. 
Therefore, all observed effects are flow-induced.
We have implemented a soft repulsion force for particles near contact.
This force becomes active when two mesh vertices of different particles come closer than one lattice constant.
However, we observed that this force is rarely required since the volume fraction is rather small and the hydrodynamic interactions are usually sufficient to repel particles at small distances.

\subsection{Simulation parameters and setup}
\label{sec_parameters}

\begin{figure}
 \centering
 \includegraphics[width=0.5\linewidth]{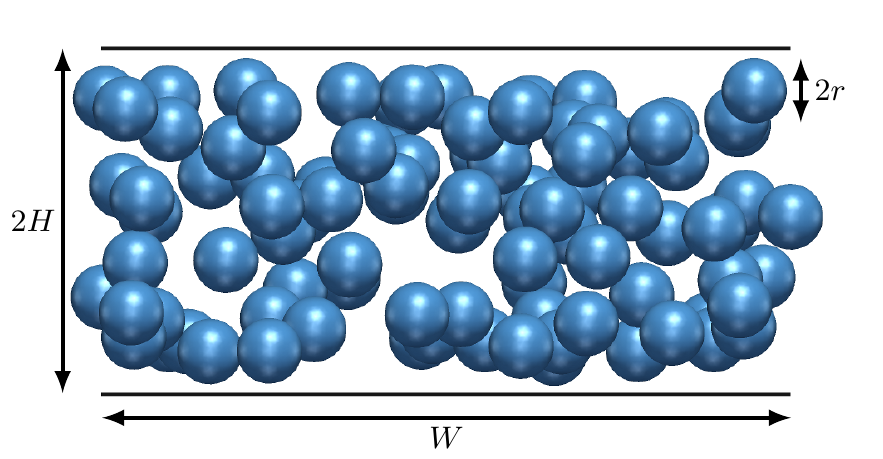}
 \caption{(Colour online) Geometrical setup. The distance between the confining walls is $2 H$, and the length and width of the channel is $W$. The undeformed particle radius is $r$. The snapshot shows a random initial state.}
 \label{fig:setup}
\end{figure}

We investigate the suspension rheology in a channel with width $W$ between two planar walls at $z = \pm H$ ($2 H$ is the distance between the walls) as illustrated in figure \ref{fig:setup}.
The system is periodic along the $x$- and $y$-axes.
Flow is induced by a constant body force density $f$ along the $x$-axis.
This geometry has been chosen for convenience; especially the data evaluation is facilitated by a planar rather than by a cylindrical or even a duct-like geometry.

Relevant simulation parameters are the volume fraction $\phi = 4 \pi N r^3 / 3 V$ (which is computed from the number $N$ of particles, their initial, undeformed radius $r$, and the total system volume $V$) and the confinement $\chi = r / H$.

In order to control the Reynolds and capillary numbers independently, three simulation parameters are available: the kinematic viscosity $\nu$ of the ambient fluid, the external force density $f$, and the particle shear modulus $\mods$.

We define the bare Reynolds number as
\begin{equation}
 \label{eq:Re_1}
 \Renumin = \frac{\hat u_0 H}{\nu} = \frac{f H^3}{2 \rho \nu^2},
\end{equation}
where $\hat u_0$ is the centreplane velocity of the unperturbed flow (\emph{i.e.}, without particles) and $\rho$ is the fluid density.
The solution for the unperturbed profile ($z = 0$ corresponds to the centreplane) is
\begin{equation}
 \label{eq:poiseuille}
 u_0(z) = \hat u_0 \left[1 - \left(\frac{z}{H}\right)^2\right], \quad \hat u_0 = \frac{f H^2}{2 \rho \nu}.
\end{equation}
The other important parameter is the shear capillary number as ratio of a typical shear stress magnitude $\bar \sigma$ to the characteristic elastic particle stress $\mods / r$:
\begin{equation}
 \label{eq:Ca_1}
 \Canum = \frac{\bar \sigma r}{\mods}.
\end{equation}
We define the characteristic shear stress $\bar \sigma$ as the average shear stress in the system.
For the unperturbed flow, the momentum balance requires the local condition
\begin{equation}
 \label{eq:stress_vs_z}
 \sigma(z) = - f z.
\end{equation}
This leads to the average stress magnitude
\begin{equation}
 \label{eq:average_stress}
 \bar \sigma := \frac{1}{2 H} \int_{-H}^{+H} \lvert \sigma(z) \rvert = \frac{f H}{2}.
\end{equation}
Equation (\ref{eq:stress_vs_z}) and (\ref{eq:average_stress}) are also valid for flows perturbed by particles, as long as the suspension is in a quasi-steady (\emph{i.e.}, on average non-accelerated) state and the stress is averaged over a sufficiently long time.

For the current study, the parameters $2 H = 60 \Delta x$ and $r = 5.9 \Delta x$ have been chosen, and the particle number is $N = 96$ for the suspension and $N = 1$ for the single-particle reference simulations.
The system size along the $x$- and $y$-axes is $W = 4 H = 120 \Delta x$ each, thus $\chi = 0.20$ is fixed and $\phi \approx 0.1$ for the suspension and $\phi \approx 0.001$ for the single particle.
Since our primary interest is not on the effect of the confinement, we choose a moderate value, avoiding both highly and weakly confined situations.

The Reynolds and capillary numbers are varied in the intervals $[3, 417]$ and $[0.003, 0.3]$, respectively (corresponding to BGK relaxation parameters $\tau \in [0.5072, 1.4]$).
Particles with $\Canum = 0.003$ are nearly rigid, and a further decrease of $\Canum$ leads to stability problems.
Capillary numbers larger than $0.3$ are problematic because the capsule deformation becomes so severe that a higher resolution would be required to capture the local curvature.
Without increasing the resolution the Reynolds number cannot be chosen much smaller than 3 or much larger than 400 due to stability and accuracy considerations.

Since there is one degree of freedom left (three control parameters to set two target parameters), we choose the parameters in a way such that the unperturbed lattice centreplane velocity $\hat u_0$ in equation (\ref{eq:poiseuille}) is $\frac{1}{30} \frac{\Delta x}{\Delta t}$.
This ensures a) small Mach numbers and therefore negligible compressibility artefacts and b) a sufficiently small time step to reduce time discretisation errors.

On the one hand, for a fixed capillary number, the Reynolds number is changed by adapting the kinematic viscosity $\nu$ in a way such that $\hat u_0$ in equation (\ref{eq:poiseuille}) remains constant.
This requires a change of the force density $f \propto \nu$ which in turn leads to a change of $\mods \propto f$ so that \Canum remains constant as can be inferred from equation (\ref{eq:Ca_1}).
On the other hand, \Canum can be changed for constant \Renum just by adapting $\mods$.
Concluding, the relevant simulation parameters can be directly obtained from the target capillary and Reynolds numbers:
\begin{equation}
 \nu = \frac{\hat u_0 H}{\Renumin}, \quad f = \frac{2 \rho \hat u_0}{\Renumin H}, \quad \mods = \frac{\rho \hat u_0 r}{\Canum \Renumin}, \quad \kappa_\alpha = 2 \mods, \quad \modb = 2.87 \cdot 10^{-3} \mods r^2,
\end{equation}
where $H = 60$, $r = 5.9$, $\hat u_0 = 1 / 30$ and $\rho = 1$ in simulation units ($\Delta x = \Delta t = 1$).

We stress that, in the present work, we focus on artificial capsules for which the selected ratio of \Canum and \Renump (where $\Renump = \dot \gamma r^2 / \nu$ is the Reynolds number on the particle scale) can be tuned. For example, healthy red blood cells (RBCs) in an aqueous environment have a fixed ratio $\Canum / \Renump = \rho \nu^2 / (\mods r) \approx 50$. Investigating inertial effects on the scale of a single RBC would require $\Renump > 1$ and therefore $\Canum > 50$ or, in physical units, $\dot \gamma > 6 \cdot 10^4 / \mathrm{s}$. For RBCs, it is therefore impossible to achieve high \Renump and small \Canum at the same time. Contrarily, by changing $r$ and $\mods$ of artificial capsules, the ratio $\Canum / \Renump$ can be controlled.

As initial condition, the undeformed spherical particles in the suspension with $\phi = 0.1$ were randomly distributed in the fluid volume avoiding overlap with each other and the walls.
The single particle was released half-way between the centreplane and one of the walls, but reference simulations with different initial positions gave the same final state.
The initial fluid density was unity everywhere, and the flow profile was fully developed according to equation (\ref{eq:poiseuille}).
Starting with a fully developed velocity profile is necessary to avoid long transients.
The time scale for acceleration of the entire fluid from a quiescent to a fully developed state is proportional to the momentum diffusion time $t_\text{md} \sim H^2 / \nu.$ Especially for high Reynolds numbers, when the kinematic viscosity $\nu$ is small, $t_\text{md}$ becomes undesirably large.

All suspension simulations ran for $6 \cdot 10^5$ time steps which turned out to be sufficient to obtain converged particle concentration profiles.
During this time, a particle at the centreplane is advected by a distance equal to about $3\,400$ particle radii.
We define the Stokes time as the time a particle at the centreplane would require to travel its own radius. Here, the Stokes time corresponds to 177 time steps.
We made sure that all single particle simulations ran until the particle reached its lateral equilibrium position.
As a rule of thumb, the required number of time steps decreased with \Renum, and the characteristic time scale for lateral migration was found to be between $50$ and $1100$ Stokes times.

\section{Results and discussion}
\label{sec_results_discussion}

We restrict ourselves to the quasi-steady-state properties of the suspensions.
\Citet{hur_sheathless_2010} provided an estimate for the channel length required for inertial particle focusing:
\begin{equation}
 \frac{L_f}{r} = \frac{\pi}{f_L \chi^3 \Renumin} \sim \frac{\mathcal{O}(10^4)}{\Renumin},
\end{equation}
where $f_L$ is a geometry-dependent parameter on the order of $0.03$ and the confinement $\chi = r / H \approx 0.2$ has been used. We therefore expect a slower convergence to a steady state for smaller $\Renumin$. In particular, for $\Renumin = 6$ we predict $L_f / r \sim \mathcal{O}(2\,000)$. Indeed, our simulation results for the suspensions show that a quasi-steady state is reached after about $1\,000$--$1\,500$ Stokes times ($\approx 1.8 \cdot 10^5$--$2.7 \cdot 10^5$ time steps), depending on the values of $\Renumin$ and \Canum.
Therefore, we drop the initial $1\,700$ Stokes times ($3 \cdot 10^5$ time steps) in all data sets and report only results obtained afterwards.
Observables are instantaneously averaged over the periodic $x$- and $y$-directions and shown as functions of the transverse coordinate $z$ only.
Any reported errors correspond to the variance due to collision-induced fluctuations about the time average.
We note that the effect of the volume fraction on the convergence time is itself an interesting problem which, however, is not within the scope of the present work.

We discuss the apparent suspension viscosity in section \ref{sec_results_viscosity} before turning our attention to the lateral particle distribution (section \ref{sec_results_particle_distribution}).
Based on these findings, we first come back to the local suspension rheology in section \ref{sec_results_local_rheology} before we describe the results obtained for a single particle in section \ref{sec_results_single}. The particle properties, such as deformation and inclination, are investigated in section \ref{sec_results_particles}.

\subsection{Apparent suspension viscosity}
\label{sec_results_viscosity}

The volume flux of a simple fluid along the $x$-axis with viscosity $\eta_0$ through a channel segment with width $W$ reads
\begin{equation}
 \label{eq:flux_simple}
 Q_0 = \int_0^W \text{d}y \int_{-H}^{+H} \text{d}z \, u_0(z) = \frac{2 f H^3 W}{3 \eta_0},
\end{equation}
where the velocity $u_0(z)$ is given by equation (\ref{eq:poiseuille}).
For a complex fluid, like a suspension, the viscosity is generally not uniquely determined in such a flow geometry, but one can use equation (\ref{eq:flux_simple}) to define an effective viscosity if the flux $Q$ is known:
\begin{equation}
 \eta := \frac{2 f H^3 W}{3 Q}.
\end{equation}
The reduced apparent viscosity is the dimensionless ratio
\begin{equation}
 \label{eq:impedance}
 \visc = \frac{\eta}{\eta_0}.
\end{equation}
For the sake of convenience, we denote $\visc$ simply as \emph{viscosity} if not clearly specified otherwise.
In terms of the unperturbed (particle-free) flux $Q_0$ and the measured flux $Q$, the viscosity is $\visc = Q_0 / Q$ and typically larger than unity.

For an observer who is not aware of the microstructure of the suspension, the presence of the particles has one major effect; the volume flux is reduced by a factor $\visc$, and the viscosity is accordingly increased by the same factor.
This means that the \emph{apparent} Reynolds number (defined by the \emph{apparent} viscosity and measured flow rate) is decreased by a factor $\visc^2$.
It is therefore convenient to use $\Renum = \Renumin / \visc^2 < \Renumin$ as the appropriate Reynolds number for the macroscopic suspension.
One should keep in mind that \Renum, contrarily to $\Renumin$, is not known \emph{a priori}.

\begin{figure}
 \centering
 \includegraphics[width=0.5\linewidth]{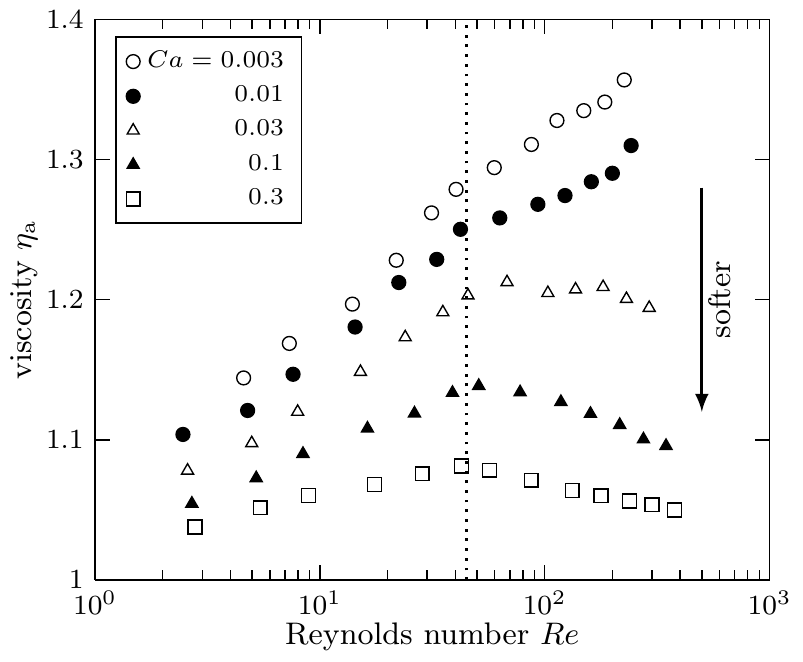}
 \caption{Viscosity $\visc$ (equation (\ref{eq:impedance})) as function of Reynolds number \Renum for different values of the capillary number \Canum for $\phi = 0.1$. Error bars are of the order of the symbol size. While the suspension exhibits a monotonic increase of $\visc$ for more rigid particles (small \Canum), the viscosity decreases for strongly deformable particles (large \Canum) for $\Renum > 45$ (dotted line). Softer particle suspensions always appear less viscous for fixed \Renum.}
 \label{fig:viscosity}
\end{figure}

Figure \ref{fig:viscosity} shows the viscosity $\visc$ of the suspension with $\phi = 0.1$ as function of \Renum for all considered capillary numbers.
One observes that, for a given value of \Renum, suspensions of softer particles always appear less viscous than suspensions of more rigid particles.
This is a well-known behaviour of deformable particle suspensions \citep{bagchi_rheology_2010}.
Soft particles adapt more easily to the flow field and disturb it less.
The particles in the small \Canum regime give rise to a monotonically increasing viscosity as function of \Renum. However, at $\Canum \geq 0.03$, the suspension shows an interesting behaviour; after an initial viscosity increase up to $\Renum \approx 45$, the viscosity decreases again.
We will explain this effect by taking into account the observed lateral particle distributions and the local rheology (section \ref{sec_results_particle_distribution} and \ref{sec_results_local_rheology}). As a remark, we expect that, in the Stokes limit ($\Renum \to 0$), the viscosity reaches a \Renum-independent plateau whose height depends only on \Canum.

\subsection{Lateral particle distribution and inertial focusing}
\label{sec_results_particle_distribution}

In order to quantify the microscopic structure of the suspension, we first focus on the lateral distribution of the particles.
Let $\phi(z)$ be the volume fraction (or particle concentration) profile between the channel walls.
We define the second moment of the particle density as a measure for the lateral distribution of the particles:
\begin{equation}
 \label{eq:M2}
 M_2 := \frac{1}{2 H} \int_{-H}^{+H} \phi(z) z^2\, \text{d}z.
\end{equation}
In order to take advantage of this quantity, we introduce the lateral displacement parameter $\Delta$ as the square root of the normalised second moment ($\Delta \propto \sqrt{M_2}$) such that a delta distribution $\phi(z) = 2 H \phi \delta (z - z_p)$ yields $\Delta = \pm \lvert z_p \rvert / H$.
This way, $\Delta$ equals the non-dimensional lateral position of a small single particle in the channel (with possible values in the interval $[-1,+1]$).
For a delta-distribution located at $z_p$ we find
\begin{equation}
 M_2 = \frac{1}{2 H} \int_{-H}^{H} 2 H \phi \delta(z - z_p) z^2\, \text{d}z = \phi z_p^2
\end{equation}
and therefore
\begin{equation}
 \label{eq:Delta}
 \Delta = \sqrt{\frac{M_2}{\phi H^2}}.
\end{equation}
The parameter $\Delta$ can be considered as the average dimensionless lateral position of the particles.
A perfectly homogeneous suspension yields $\Delta = \sqrt{1/3}$.
Additionally to $\Delta$, we define the reduced instantaneous depletion layer thickness $\depletion$ as the minimum distance of any particle's surface element to any confining wall, normalised by $H$.

\begin{figure*}
 \subfloat[\label{fig:Delta} lateral displacement parameter]{\includegraphics[width=0.475\linewidth]{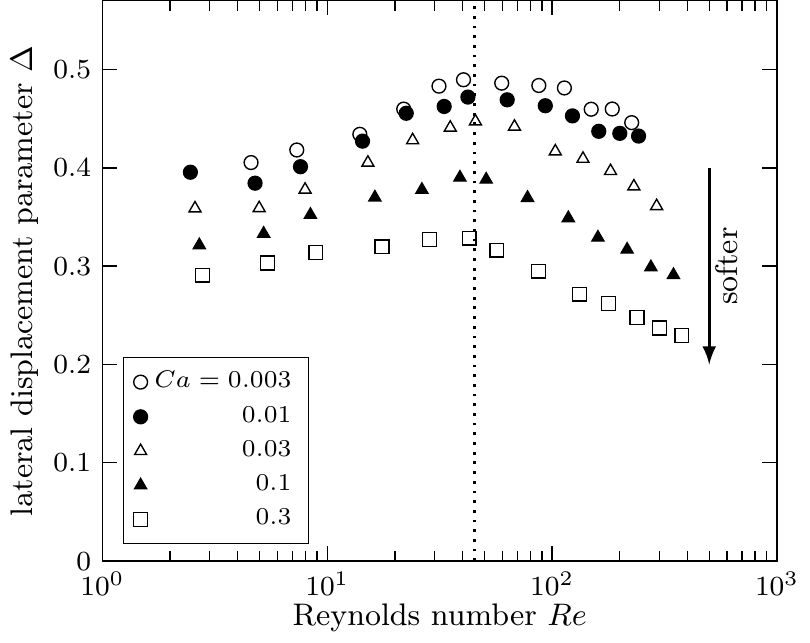}} \hfill
 \subfloat[\label{fig:delta} depletion layer thickness]{\includegraphics[width=0.475\linewidth]{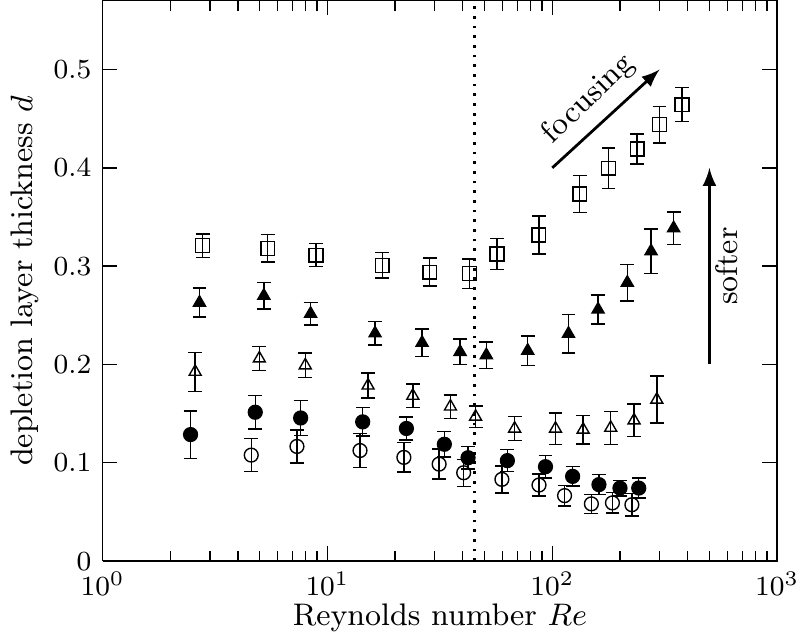}}
 \caption{Characterisation of the lateral particle distribution for $\phi = 0.1$. (a) Lateral displacement parameter $\Delta$ (equation (\ref{eq:Delta})). Errors are of the order of the symbol size. A perfectly homogeneous suspension corresponds to a value of $1 / \sqrt{3} \approx 0.57$ (top edge). Smaller values indicate a concentration of particles closer to the centreplane. On average, softer particles (larger \Canum) are always closer to the centreplane for a given \Renum. Note the presence of a maximum for each curve at $\Renum \approx 45$ (dotted line). Beyond this value, particles tend to move closer to the centreplane again. (b) Reduced depletion layer thickness $\depletion$. For $\depletion = 0$, particles would touch the walls; $\depletion = 1$ corresponds to the centreplane. The legend is valid for both plots. The effect of inertia is almost negligible for $\Renum < 10$. More rigid particles (smaller \Canum) tend to form the thinnest depletion layer. Deformable particles show a strongly \Renum-dependent behaviour; for $\Renum > 45$ (dotted line), the depletion layer grows significantly.}
 \label{fig:Delta_delta}
\end{figure*}

The lateral displacement parameter $\Delta$ as function of \Renum is shown for different $\Canum$ in figure \ref{fig:Delta}.
Softer particles are always more strongly concentrated near the centreplane for fixed \Renum.
For all values of \Canum, $\Delta$ has a maximum in the vicinity of $\Renum = 45$.
Except for their magnitude, the shapes of the $\Delta$-curves are rather similar for different capillary numbers.
Interestingly, \Citet{shin_inertial_2011} also noticed a maximum lateral displacement of a single 2D elastic capsule in inertial flow for $\Renum = 30$ in a channel with a confinement of $0.1$.

\begin{figure}
 \centering
 \subfloat[\label{fig:snapshot_re006_ca0003} $\Renumin = 6$, $\Canum = 0.003$]{\includegraphics[height=1.8cm]{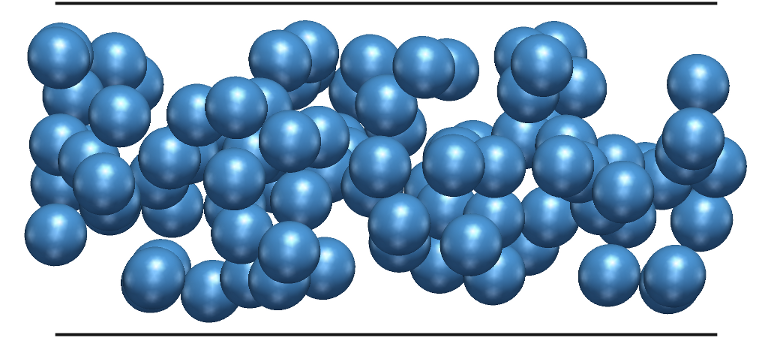} \includegraphics[height=1.8cm]{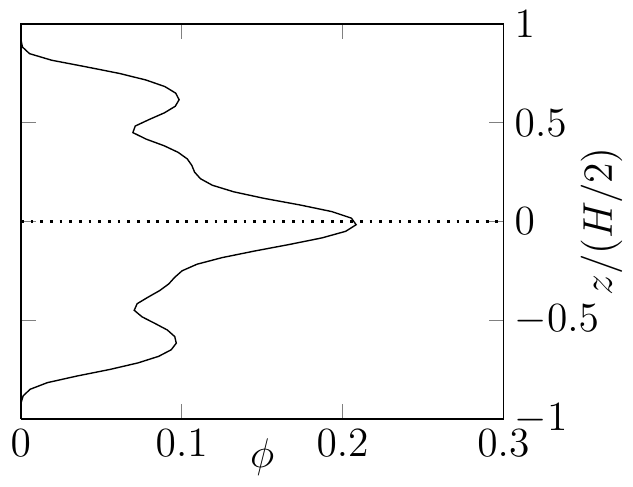}} \hfill
 \subfloat[$\Renumin = 6$, $\Canum = 0.3$]{\includegraphics[height=1.8cm]{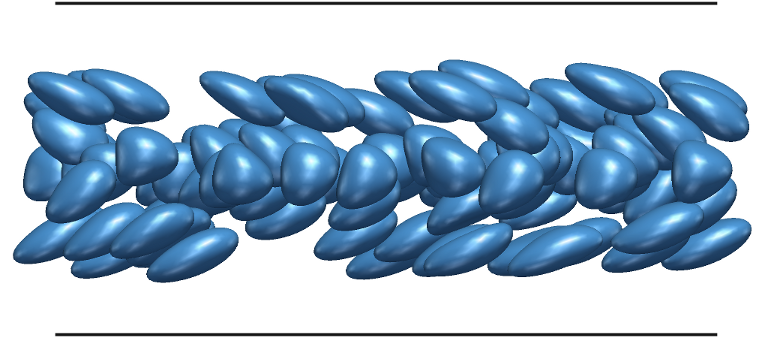} \includegraphics[height=1.8cm]{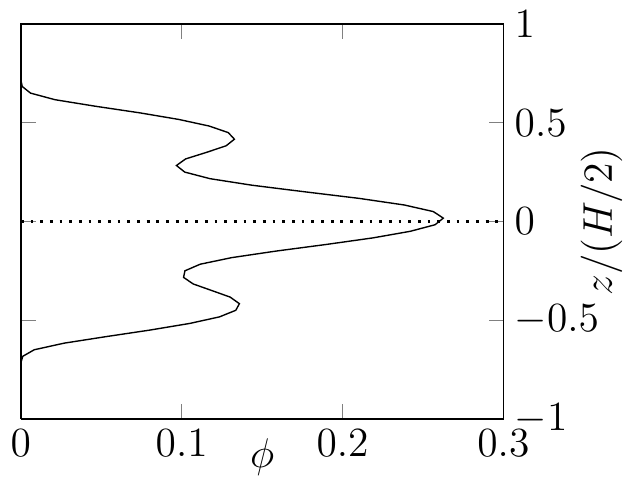}} \\
 \subfloat[$\Renumin = 50$, $\Canum = 0.003$]{\includegraphics[height=1.8cm]{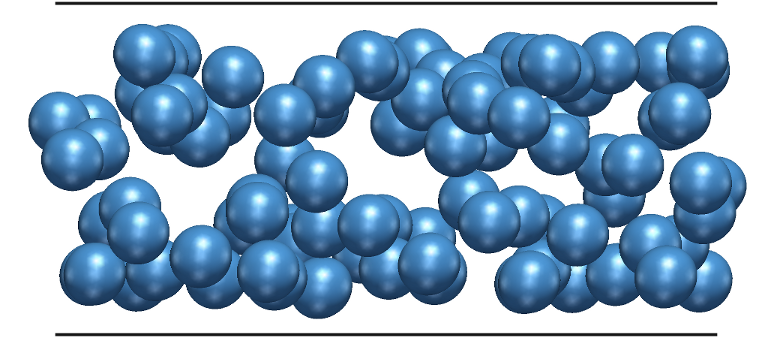} \includegraphics[height=1.8cm]{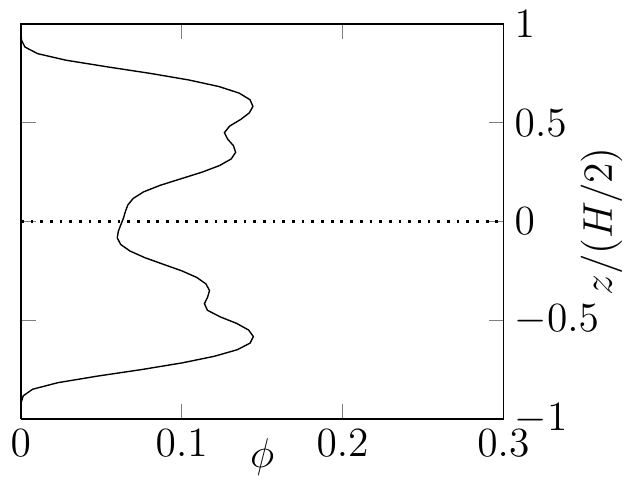}} \hfill
 \subfloat[$\Renumin = 50$, $\Canum = 0.3$]{\includegraphics[height=1.8cm]{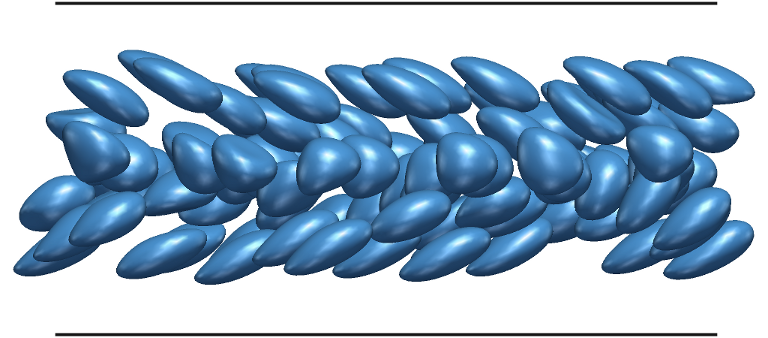} \includegraphics[height=1.8cm]{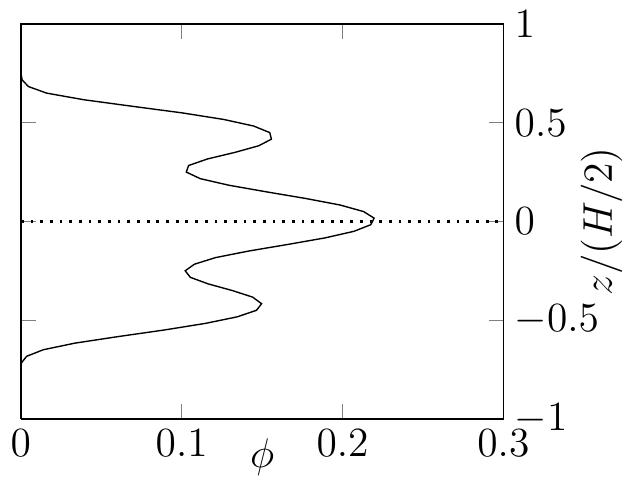}} \\
 \subfloat[$\Renumin = 417$, $\Canum = 0.003$]{\includegraphics[height=1.8cm]{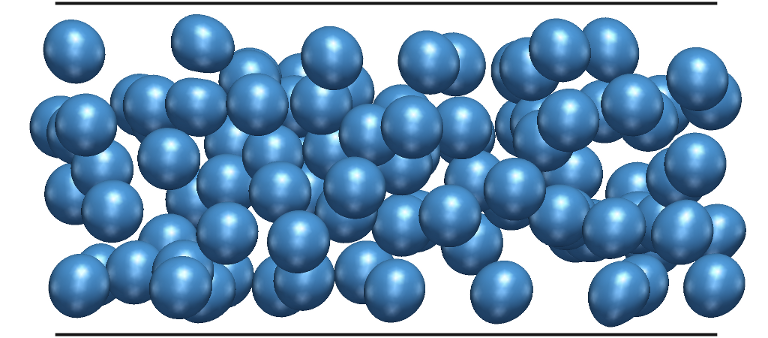} \includegraphics[height=1.8cm]{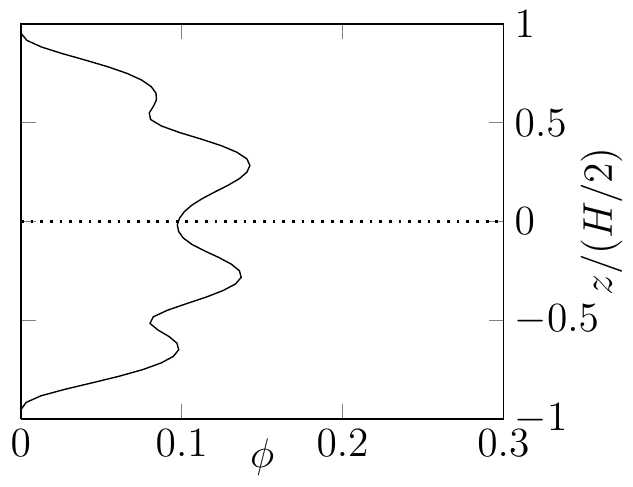}} \hfill
 \subfloat[$\Renumin = 417$, $\Canum = 0.3$]{\includegraphics[height=1.8cm]{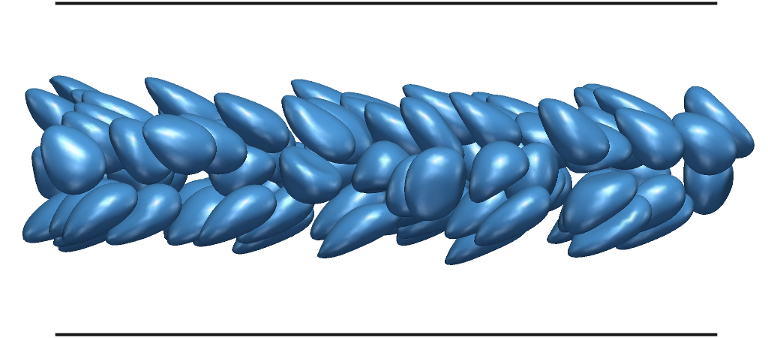} \includegraphics[height=1.8cm]{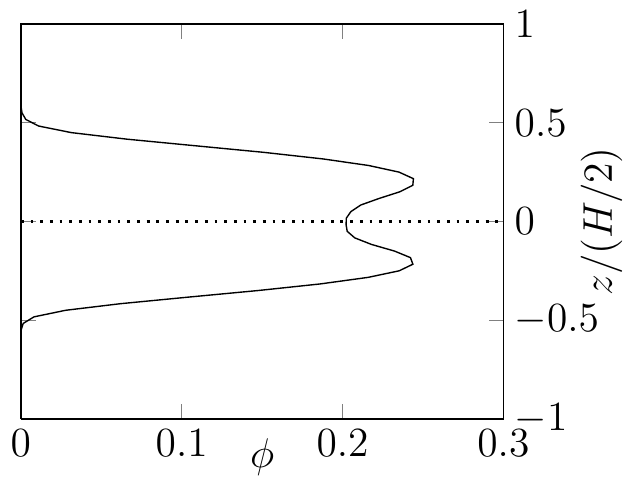}}
 \caption{\label{fig:snapshots_density} (Colour online) Exemplary instantaneous quasi-steady-state configurations of the particles (snapshots) and corresponding time-averaged particle concentration profiles (plots) for different combinations of Reynolds and capillary numbers. The dotted lines correspond to the centreplane. (a) For $\Renumin = 6$ and $\Canum = 0.003$ neither inertial nor deformability effects are very important. (b) Increasing the capillary number to $0.3$ leads to a migration of the particles towards the centreplane. (c) For $\Renumin = 50$ and $\Canum = 0.003$ the nearly rigid particles are pushed towards the walls and prefer lateral positions resembling the Segré-Silberberg effect. (d) This behaviour is distorted when the particles are strongly deformable ($\Canum = 0.3$), and the concentration profile is rather similar to that for smaller $\Renumin$ in (b). (e) For large Reynolds numbers ($\Renumin = 417$) the Segré-Silberberg-like behaviour of the nearly rigid particles ($\Canum = 0.003$) is less pronounced, and the particles are closer to the centreplane on average. (f) When both inertial effects and deformability are important ($\Renumin = 417$, $\Canum = 0.300$), the particles are strongly focused and concentrated in the central region.}
\end{figure}

Figure \ref{fig:delta} contains the data for the depletion layer thickness $\depletion$.
More rigid particles tend to be closer to the confining walls for increasing \Renum.
For the stiffest particles ($\Canum = 0.003$), the depletion layer thickness at small \Renum is about $10\%$ of the channel half-width, and it decreases to about $5\%$ for the largest simulated Reynolds number.
Already for $\Canum = 0.03$, but especially for even larger \Canum, the function $\depletion(\Renum)$ has a minimum.
For $\Canum = 0.3$, $\depletion$ is almost constant up to $\Renum \approx 40$.
Beyond, the depletion layer significantly increases and reaches values of nearly $50\%$ for $\Renum = 417$.
The particles are strongly focused towards the centreplane upon increasing \Renum;
we therefore call this effect \emph{inertial focusing}. It should be noted that this term has been used by other authors to describe different effects \citep{di_carlo_inertial_2009, humphry_axial_2010, martel_inertial_2012}.
By comparing figure \ref{fig:viscosity} and figure \ref{fig:delta}, we also observe that the thickness of the depletion layer and the suspension viscosity are tightly connected. The viscosity increases when the depletion layer thickness decreases and vice versa. A large depletion layer thickness $\depletion$ reduces the viscous dissipation in wall vicinity. This is related to the Fåhræus-Lindqvist effect \citep{fahraeus_viscosity_1931} observed in blood flow; however, in the present case, the depletion layer thickness is controlled by the Reynolds number and not the degree of confinement of red blood cells in a vessel.

Interestingly, the peak in $\Delta(\Renum)$ and the increase of the depletion layer $\depletion(\Renum)$ for large \Canum (figure \ref{fig:Delta_delta}) are observed at a similar Reynolds number ($\approx 45$) where also the kink in the apparent viscosity $\visc(\Renum)$ in figure \ref{fig:viscosity} is located.
This suggests a strong dependence of the viscosity on the particle distribution.
We will come back to this observation in section \ref{sec_results_local_rheology}.
Similarly to the data in figure \ref{fig:viscosity} it is expected that all curves in figure \ref{fig:Delta_delta} reach a \Canum-dependent plateau for $\Renum \to 0$.

Due to the non-homogeneous stress and particle density distributions it is difficult to achieve data collapse in figure \ref{fig:Delta_delta} and to find simple, globally defined scaling laws.

Some characteristic particle configurations and density profiles $\phi(z)$ are shown in figure \ref{fig:snapshots_density}.
In the limit of small Reynolds and capillary numbers ($\Renumin = 6$ and $\Canum = 0.003$), the particles are basically undeformed and unevenly distributed throughout the channel.
By increasing their deformability ($\Canum = 0.3$), the particles move closer towards the centreplane. 
At the intermediate Reynolds number $\Renumin = 50$ the nearly rigid particles
($\Canum = 0.003$) show an affinity for distances half-way between the
centreplane and the walls which reminds of a Segré-Silberberg-like behaviour
\citep{segre_behaviour_1962, segre_behaviour_1962-1}.
This effect disappears for $\Canum = 0.3$ where configuration and particle distribution look very similarly to that at $\Renumin = 6$ and $\Canum = 0.3$.
We therefore conclude that the Segré-Silberberg effect can be suppressed by choosing sufficiently soft particles.

For even larger $\Renumin$ ($\Renumin = 417$) the more rigid particles are more homogeneously distributed, but they show a strong inertial focusing when both \Renum and \Canum are large.
In the latter case, the depletion layer is very pronounced and the average particle concentration near the centreplane is greater than $20\%$ and therefore more than twice as large as the overall average.
In all cases, several individual density peaks are visible.
These are probably due to the relatively large confinement ($\chi = 0.2$) \citep{li_wall-bounded_2000, zurita-gotor_layering_2012}.

Before we turn our attention more closely to the mechanisms responsible for the lateral particle distribution in section \ref{sec_results_single}, let us first discuss the local suspension rheology.

\subsection{Local rheology and dependence of viscosity on suspension microstructure}
\label{sec_results_local_rheology}

\begin{figure}
 \centering
 \includegraphics[width=0.5\linewidth]{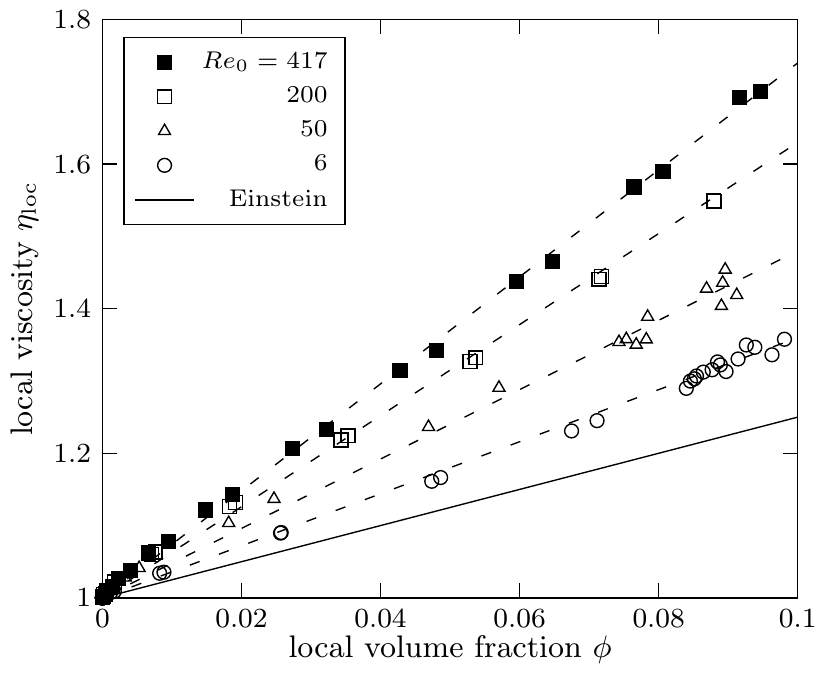}
 \caption{The local viscosity is shown as function of the local volume fraction for $\Canum = 0.03$ and a few Reynolds numbers. The viscosity is a monotonically increasing function of both the volume fraction $\phi$ and the Reynolds number \Renum. For the sake of comparison, we also show Einstein's viscosity ($1 + \frac{5}{2} \phi$) one would expect for a dilute suspension of rigid spheres in a simple viscous shear flow. The local viscosity is roughly linear in $\phi$ as indicated by the dashed lines as guide for the eyes.}
 \label{fig:viscosity_vs_density}
\end{figure}

In order to quantify the \emph{local} rheology, we define the reduced local viscosity $\viscloc$ in a plane at a given $z$-value as the ratio of the known stress at position $z$ (equation (\ref{eq:stress_vs_z})) divided by the measured average shear rate $\dot \gamma(z)$ in this plane:
\begin{equation}
 \label{eq:local_viscosity}
 \viscloc(z) := \frac{1}{\eta_0} \frac{\sigma(z)}{\dot \gamma(z)}.
\end{equation}
Figure \ref{fig:viscosity_vs_density} shows the dependence of $\viscloc$ on the local volume fraction for a few selected parameter sets for $\Canum = 0.03$ (similar results have been obtained for the other capillary numbers).
We construct these data sets from a combination of $\phi(z)$ (figure \ref{fig:snapshots_density}) and $\viscloc(z)$ (not shown).
All data in the central region ($\lvert z / r \rvert < 1$) are excluded due to large fluctuations; both the stress and the shear rate are small for $z \to 0$, and their ratio is prone to significant noise.

As expected, the local viscosity is a monotonically increasing function of the local volume fraction for all investigated data sets.
The $\phi$-dependence is roughly linear up to $\phi = 0.1$.
Furthermore, local viscosities in our simulations are generally decreasing with \Canum as the particles become softer and contribute less to dissipation (data not shown).

We also notice that $\viscloc$ is strictly increasing with \Renum for fixed $\phi$. This behaviour will be explained based on the particle properties in section \ref{sec_results_particles}.
The viscosity curves $\viscloc(\phi)$ shown in figure \ref{fig:viscosity_vs_density} are larger than the corresponding Einstein viscosity ($1 + \frac{5}{2} \phi$) of a dilute suspension of hard spheres with negligible inertia.
It is not clear how $\viscloc(\phi)$ behaves for $\Renum \to 0$ since this parameter region is not accessible in the current simulation setup.

\begin{figure}
 \centering
 \includegraphics[width=0.7\linewidth]{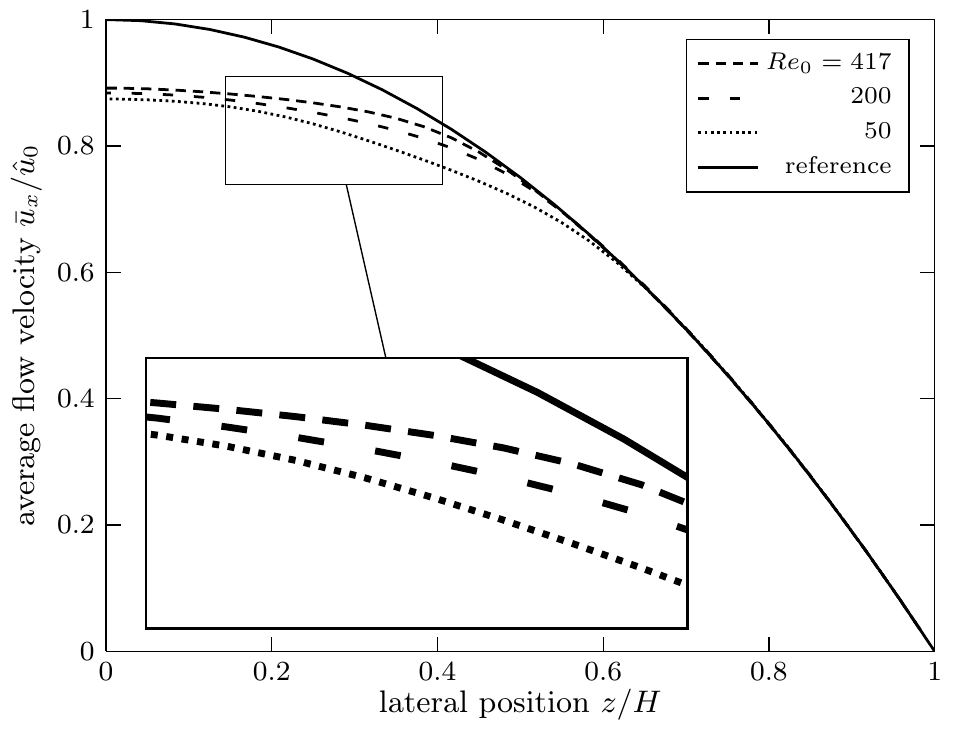}
 \caption{Representative averaged velocity profiles $\bar u_x(z)$ for $\Canum = 0.3$ and $\phi = 0.1$. The reference profile equation (\ref{eq:poiseuille}) without particles is shown as solid line. For $\Renum = 417$, the area below the curve (\emph{i.e.}, the total flux) is larger than for $\Renum = 50$ or $200$.}
 \label{fig:velocity_profiles}
\end{figure}

We can now turn our attention back to the apparent viscosity $\visc$ (section \ref{sec_results_viscosity}). There are two mechanisms which lead to a variation of $\visc(\Renum)$.
\begin{enumerate}
 \item For a fixed density distribution $\phi(z)$, the viscosity $\visc$ is expected to grow with \Renum because $\viscloc$ increases with \Renum (figure \ref{fig:viscosity_vs_density}).
 \item A \Renum-induced redistribution of particles towards the centreplane tends to reduce $\visc$ because less particles are located in high-dissipation regions. Viscous dissipation generally scales like the product of shear rate $\dot \gamma$ and fluid shear stress $\sigma_\text{f}$. As a consequence, most of the dissipation in a Poiseuille-like flow occurs in the wall region where $\dot \gamma$ and $\sigma_\text{f}$ are large. Therefore, $\visc$ is dominated by the contribution of $\viscloc$ in wall vicinity, whereas $\viscloc$ in the central region is less relevant.
\end{enumerate}
The first effect can be seen for suspensions with $\Canum = 0.003$ and $0.01$; figure \ref{fig:Delta_delta} reveals that the lateral particle distribution changes only slightly with \Renum for $\Renum > 45$. Yet, $\visc$ shows a strong increase (figure \ref{fig:viscosity}), which can be attributed to the increase of $\viscloc$ with \Renum.
The second effect is particularly important for $\Canum = 0.1$ and $0.3$; the increase of $\viscloc$ with \Renum is overcompensated by the strong focusing of particles (figure \ref{fig:Delta_delta}), and $\visc$ decreases (figure \ref{fig:viscosity}).
This can also be seen in figure \ref{fig:velocity_profiles} where a few selected velocity profiles for $\Canum = 0.3$ are shown.
For $\Renum > 45$, particles are more concentrated near the centreplane, and the velocity profiles approach the reference profile without particles.
This results in a larger flux $Q$, and $\visc$ decreases.
For $\Canum = 0.03$, both above-mentioned effects basically compensate each other, and $\visc$ remains roughly constant for $\Renum > 45$ (figure \ref{fig:viscosity}).
In general, the velocity profiles in figure \ref{fig:velocity_profiles} show a pronounced flattening near the centreplane which is typical for shear-thinning fluids.

Concluding we note that, although the suspensions investigated here show a
strict increase of $\viscloc(\Renum)$ when observed locally, $\visc$ can
decrease since the particles are inhomogeneously distributed throughout the
channel and this microstructure changes with \Renum. Such an effect would
therefore not be observed in an unbounded simple shear flow where the stress
and dissipation are homogeneous on average.

\subsection{Mechanisms responsible for lateral particle distribution}
\label{sec_results_single}

In a steady state, the lateral position of particles (either isolated or in a denser suspension) is the consequence of the balance of forces pointing towards and away from the centreplane.
For rigid particles at intermediate Reynolds numbers, the inertial lift is balanced by the wall repulsion, which gives rise to the Segré-Silberberg effect.
Deformable particles tend to approach the centreplane due to a combination of wall repulsion and deformability in the presence of a curved velocity profile \citep{kaoui_lateral_2009}.
This effect may be partially compensated by an inertial lift force if the Reynolds number is sufficiently large.
In a non-dilute suspension, not all particles can be located at the centreplane at the same time.
Hydrodynamic interactions usually lead to non-reversible particle trajectories and a diffusive motion \citep{eckstein_self-diffusion_1977, pranay_depletion_2012} which result in a wide distribution of particles across the channel.
However, \citet{doddi_effect_2008} reported that the nature of the hydrodynamic interactions of deformable capsules changes in the presence of inertia.
The diffusive motion of two colliding capsules in shear flow can be replaced by spiralling motion at larger \Renum.
In a suspension of sufficiently soft particles, this may lead to a reduced hydrodynamic repulsion and therefore to a larger concentration near the centreplane as we observe it in figure \ref{fig:Delta_delta} and \ref{fig:snapshots_density}.

Therefore, we have investigated two distinct ingredients responsible for the final lateral distribution of the suspended particles:
\begin{enumerate}
 \item What is the lateral equilibrium position of a single particle ($\phi \to 0$)?
 \item How do particle-particle collisions in the suspension ($\phi = 0.1$) contribute?
\end{enumerate}

\begin{figure}
 \centering
 \includegraphics[width=0.5\linewidth]{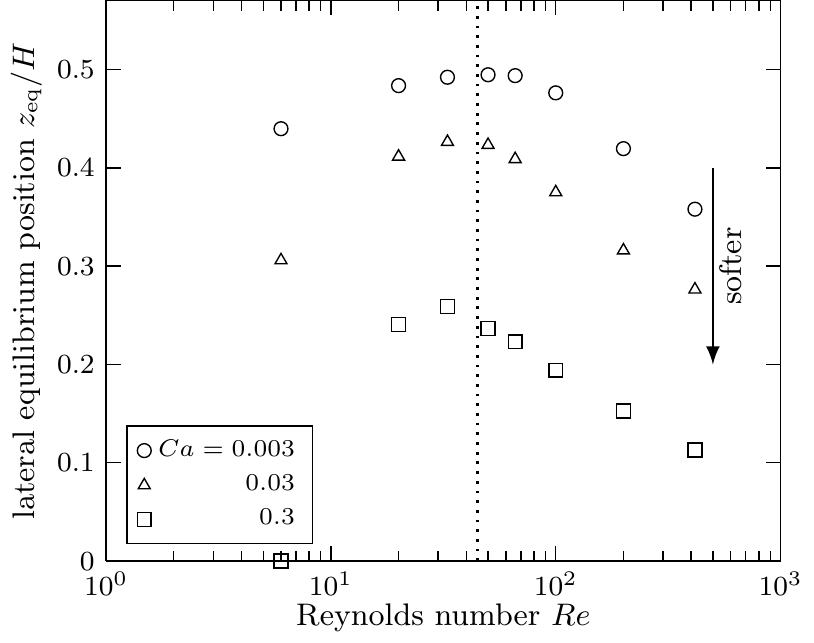}
 \caption{Lateral equilibrium position $z_\mathrm{eq}$ of a single capsule ($N = 1$, $\phi = 0.001$) as function of \Renum for different values of \Canum. Note the data point $z_\mathrm{eq} = 0$ for $\Renum = 6$ and $\Canum = 0.3$. The vertical dotted line marks $\Renum = 45$ (\emph{cf.}~figure \ref{fig:Delta_delta}).}
 \label{fig:lateral_single}
\end{figure}

Figure \ref{fig:lateral_single} shows the lateral equilibrium position $z_\mathrm{eq}$ of a single capsule ($N = 1$, $\phi = 0.001$) in an otherwise identical environment as the suspension with $\phi = 0.1$.
The similarity to the lateral displacement parameter $\Delta$ in figure \ref{fig:Delta} is striking.
In particular, the curves $z_\mathrm{eq}(\Renum)$ for different \Canum feature a peak near $\Renum = 45$ at which the particle is closest to one of the walls.
This leads to the assumption that the overall behaviour of the lateral particle distribution in the suspension, and therefore the inertial focusing, is strongly determined by that of a single particle.
We note that for $\Renum \to 0$, the capsule should eventually reach the centreplane for any finite value of $\Canum$.
Therefore, inertia is responsible for the finite values of $z_\mathrm{eq}$.
However, for increasing \Canum, the deformability effect becomes more important and the capsule reaches a configuration closer to the centreplane.
Interestingly, we observe only the capsule for $\Renum = 6$ and $\Canum = 0.3$ reaching the centreplane.
For this particular data point, deformability effects are strongest and inertia effects are weakest.
\cite{shin_inertial_2011} observed a similar variation of $z_\mathrm{eq}(\Renum)$ for simulated deformable capsules in 2D.
In particular, for a confinement of $\chi = 0.2$, they observed the maximum of $z_\mathrm{eq}(\Renum)$ between $\Renum = 30$ and $40$, depending on the details of the particle deformability (note that their definition of \Renum is identical to ours).

\begin{figure*}
 \subfloat[\label{fig:vel_fluct1} $\Canum = 0.03$]{\includegraphics[width=0.475\linewidth]{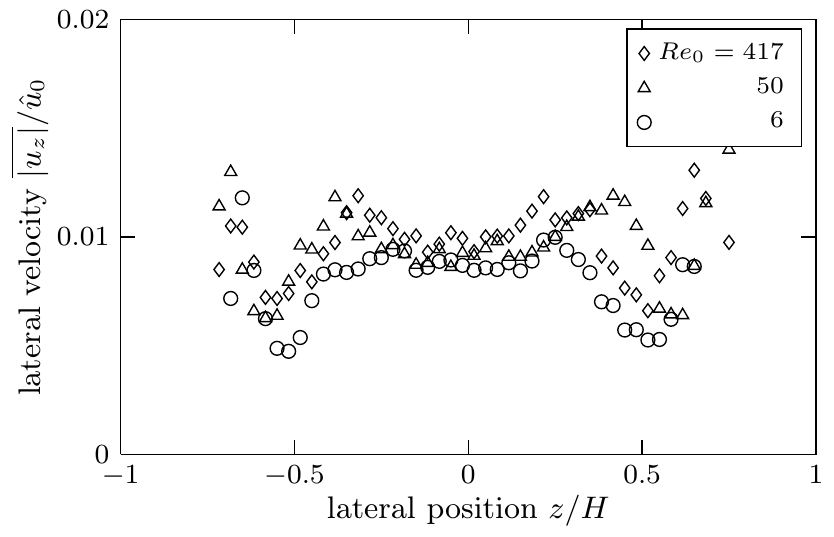}} \hfill
 \subfloat[\label{fig:vel_fluct2} $\Canum = 0.3$]{\includegraphics[width=0.475\linewidth]{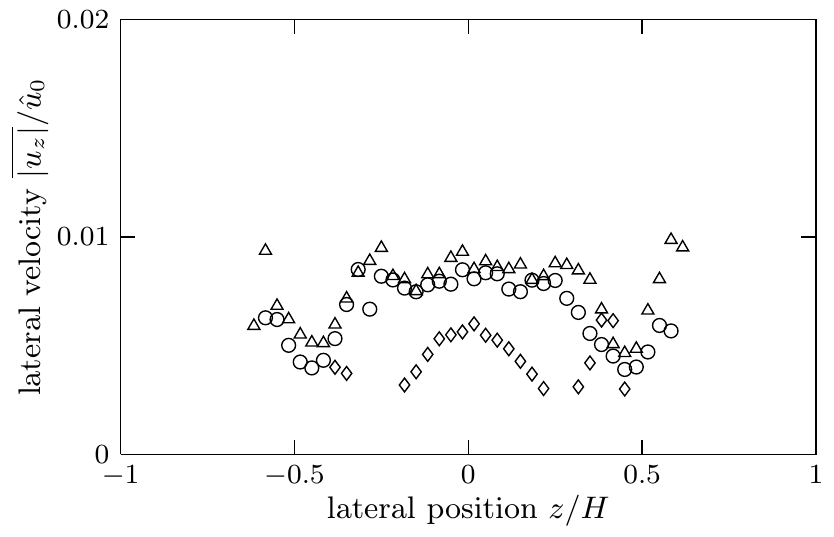}}
 \caption{Average of the lateral particle velocity magnitude $\lvert u_z \rvert$ normalised by the reference velocity $\hat u_0$ for (a) $\Canum = 0.03$ and (b) $\Canum = 0.3$. The legend is valid for both plots. Particular attention should be paid to the fluctuations for $\Renumin = 417$ and $\Canum = 0.3$ ($\diamond$ in (b)): Velocity fluctuations are significantly smaller than for smaller values of $\Renumin$ at the same \Canum, which is not observed for $\Canum = 0.03$. This suggests that the combination of large \Renum and \Canum reduces the particle dispersion which facilitates particle focusing.}
 \label{fig:vel_fluct}
\end{figure*}

The lateral particle distributions as shown in figure \ref{fig:Delta_delta} and \ref{fig:snapshots_density} are not only a result of the single particle behaviour in figure \ref{fig:lateral_single}.
If this was the case, one would find all particles located on two planes corresponding to $\pm z_\mathrm{eq}$.
This is obviously not the case, so particle-particle interactions play an important role as well.
For example, we find $z_\mathrm{eq}(\Renum = 417) / H = 0.36$, $0.28$ and $0.11$ for $\Canum = 0.003$, $0.03$ and $0.3$, respectively, but $\Delta = 0.45$, $0.36$, $0.23$ for the suspension for the same values of \Renum and \Canum.
Particle-particle interactions therefore lead to a \emph{dispersion} of the particles, pushing them farther towards the walls.
Figure \ref{fig:vel_fluct} shows the average of the magnitude of the lateral particle velocity, $\overline{\lvert u_z\rvert} / \hat u_0$, which is an indicator for lateral particle dispersion.
We observe that the fluctuation profile for $\Canum = 0.03$ does not strongly depend on the Reynolds number (figure \ref{fig:vel_fluct1}).
However, a combination of large \Canum and \Renum leads to a dramatic decrease of the lateral particle motion (figure \ref{fig:vel_fluct2}) which could be related to the effect described by \citet{doddi_effect_2008} and further facilitates inward migration and therefore particle focusing.

In the present case, it is expected that (i) the wall repulsion, (ii) the particle deformation in the presence of a curved velocity profile, (iii) the inertial lift and (iv) the hydrodynamic dispersion of particles all play a role.
These contributions are generally different functions of \Canum and \Renum and also of volume fraction and confinement.

\subsection{Particle properties}
\label{sec_results_particles}

After having presented the suspension properties, let us turn to the behaviour of the suspended particles.
All particles have been observed to perform tank-treading-like dynamics at all times. This is in agreement with previous investigations of initially spherical capsules \citep{yazdani_tank-treading_2011}. We note that initially non-spherical capsules can also undergo tumbling motion \citep{shin_dynamics_2012}.
Only the single particle on the centreplane ($\Renum = 6$, $\Canum = 0.3$) does not perform tank-treading, but rather assumes a steady bullet-shape configuration.
In the suspension, deformations and inclinations are fluctuating due to particle-particle interactions.

The particle deformation is quantified by the Taylor deformation parameter
\begin{equation}
 D = \frac{a - c}{a + c} \geq 0
\end{equation}
where $a$ and $c$ are the largest and smallest axes of the inertia ellipsoid of the deformed particle \citep{kruger_efficient_2011}.
Undeformed particles fulfil $a = c$ and therefore $D = 0$.
The deformation is related to the constitutive law of the particles and the fluid stresses on the particle surface which are responsible for their deformation.
Based on the analytical work by \citet{barthes-biesel_time-dependent_1981}, one can compute the deformation parameter of a capsule in simple shear flow for a given capillary number in the small deformation regime (\emph{i.e.} for particle shapes with a small deviation from that of a sphere).
We consider deformations large if $D > 0.1$ (corresponding to an aspect ratio of $a / c > 1.22$).
For the elasticity model employed in the current work, the relation between the deformation parameter and the capillary number reads
\begin{equation}
 \label{eq:deformation_analytical}
 D = \frac{45}{4} \frac{\frac{\moda}{\mods} + \frac{2}{3}}{2 \frac{\moda}{\mods} + 1} \Canum
\end{equation}
up to first order in \Canum.
As $\moda / \mods = 2$ is fixed in the present study, one predicts $D = 6\, \Canum$ in the small deformation limit.
It has to be noted that equation \eqref{eq:deformation_analytical} is only valid for $\modb = 0$ which is not the case in our simulations.
However, as $\modb$ is rather small, it is expected that the bending elasticity becomes only important when the local membrane curvature is large.
Due to equation (\ref{eq:stress_vs_z}) the shear stress, and therefore the local capillary number, increases with distance from the centreplane (\emph{i.e.}, $\Canum = \Canum(z)$), and the particle deformation is an increasing function of the lateral particle position $z$.

\begin{figure*}
 \centering
 \subfloat[deformation ($\Canum = 0.003$)]{\includegraphics[width=0.48\linewidth]{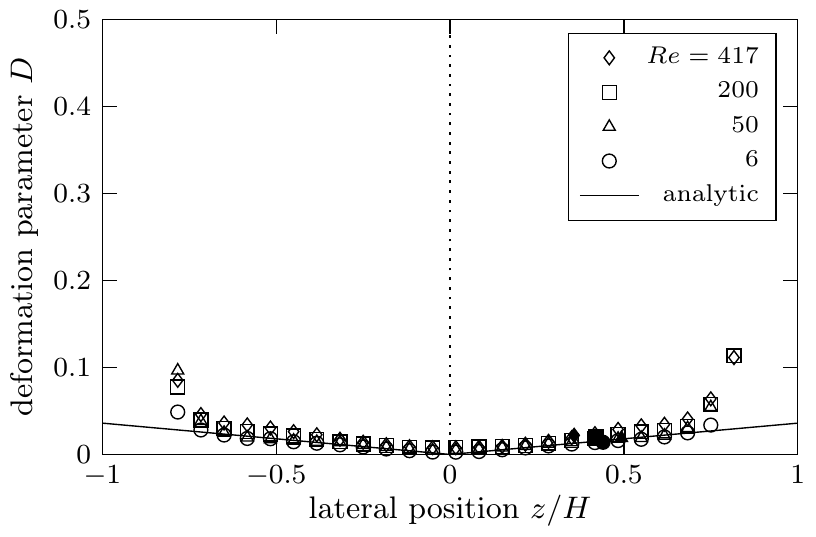}} \hfill
 \subfloat[inclination ($\Canum = 0.003$)]{\includegraphics[width=0.48\linewidth]{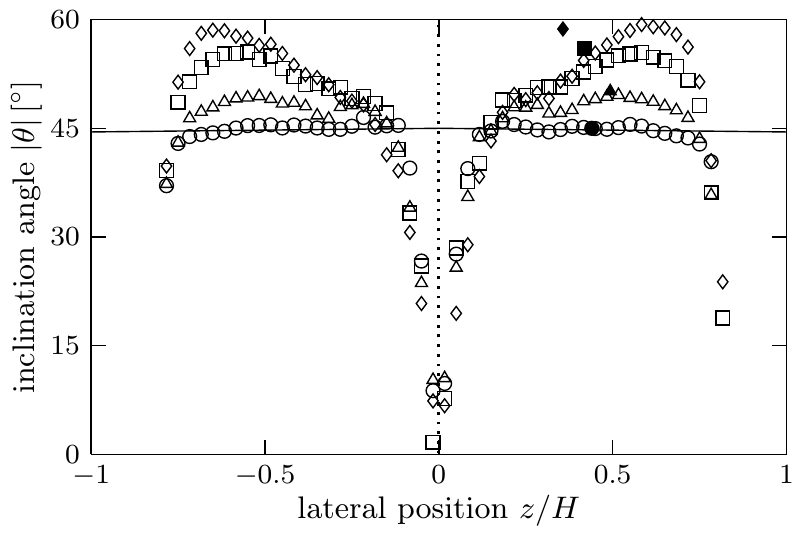}} \\
 \subfloat[deformation ($\Canum = 0.03$)]{\includegraphics[width=0.48\linewidth]{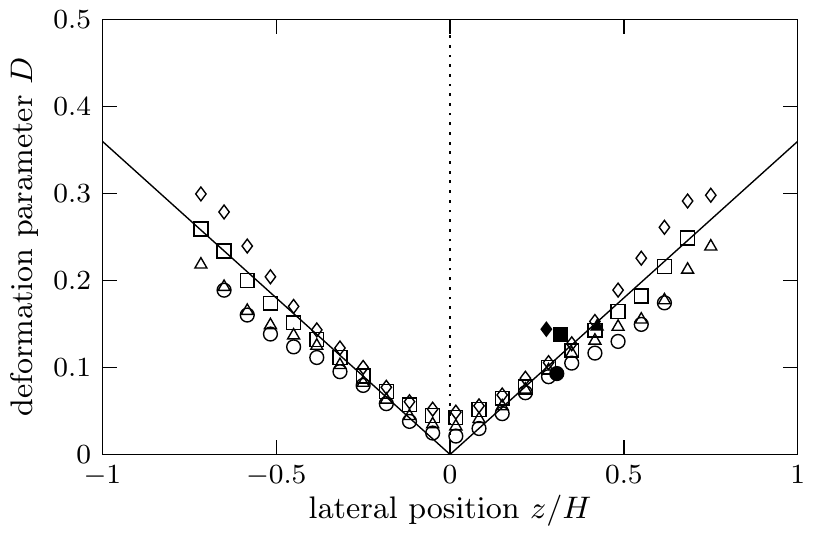}} \hfill
 \subfloat[inclination ($\Canum = 0.03$)]{\includegraphics[width=0.48\linewidth]{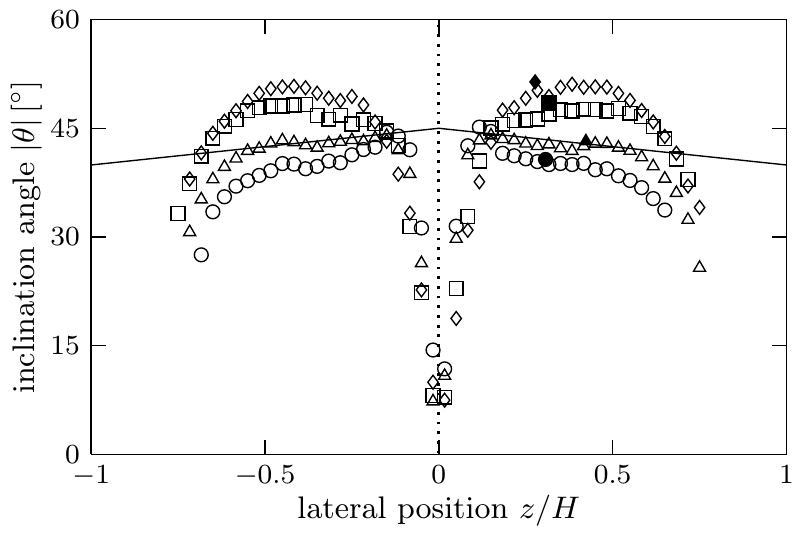}} \\
 \subfloat[deformation ($\Canum = 0.3$)]{\includegraphics[width=0.48\linewidth]{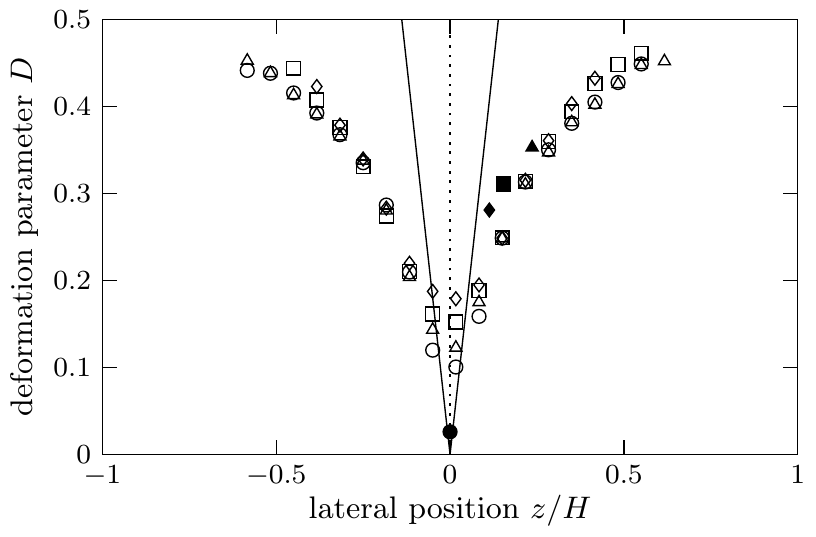}} \hfill
 \subfloat[inclination ($\Canum = 0.3$)]{\includegraphics[width=0.48\linewidth]{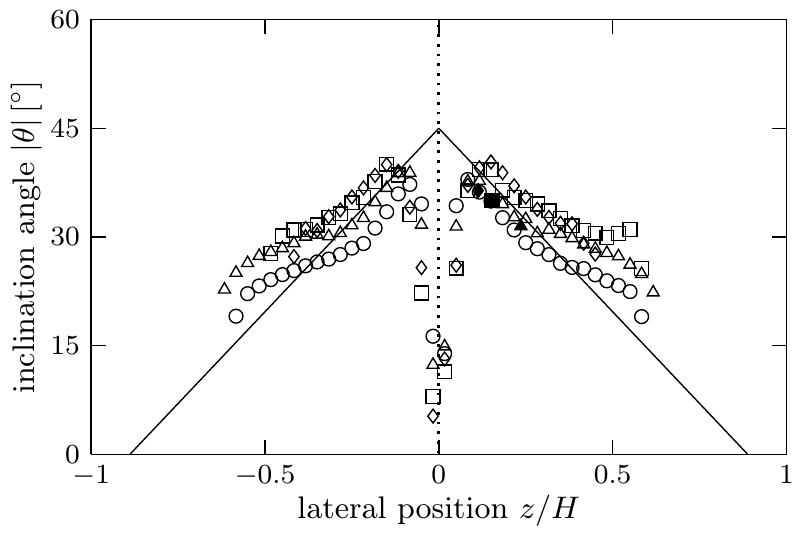}} \\
 \caption{Deformation parameter $D$ (left column) and inclination angle $\theta$ (right column) for different $\Renum$ and (a), (b) $\Canum = 0.003$, (c), (d) $\Canum = 0.03$ and (e), (f) $\Canum = 0.3$. The legend is valid for all plots. For the suspension (open symbols), only every second data point of $D$ is shown for the sake of clarity. Filled symbols denote the data points for a single particle. The vertical dotted line corresponds to the centreplane. The inclined solid lines correspond to the small deformation predictions (equation \eqref{eq:deformation_analytical} \citep{barthes-biesel_time-dependent_1981} and \eqref{eq:inclination_analytical} \citep{barthes-biesel_motion_1980}, respectively). The small deformation theory predicts the results for small \Canum and \Renum well. Inertia effects tend to increase the particle deformation and inclination.}
 \label{fig:deformation+inclination_profiles}
\end{figure*}

The time-averaged particle deformation parameter $D(z)$ is presented in the left column of figure \ref{fig:deformation+inclination_profiles} together with the data points for a single particle in its final equilibrium state.
Equation (\ref{eq:deformation_analytical}) is shown in the plots as a solid black line.
For the smallest capillary number ($\Canum = 0.003$), all particles are only slightly deformed.
The deformations at $\Renum = 6$ match the theory for single capsules in simple shear flow except for particles which are closest to the walls.
The reason is that the flow is inhomogeneous and the particles are extended.
The outermost particles are in very close proximity to the walls (figure \ref{fig:snapshot_re006_ca0003}); this leads to an additional particle deformation.
We emphasise that the small deformation theory is only strictly valid for single isolated particles in simple shear flow.
Still, it provides a good approximation in the present case.

For $\Canum = 0.03$, where the deformation is larger and therefore less well described by the linear relation in equation \eqref{eq:deformation_analytical}, we observe that particle deformation increases with the Reynolds number.
A similar trend has been observed for simulated three-dimensional capsules \citep{doddi_effect_2008} and two-dimensional vesicles \citep{kim_numerical_2012, laadhari_vesicle_2012}.
This can be understood by considering the local particle reference frame; for a tank-treading capsule or vesicle, the fluid circulates around the particle.
The curvature radius of the streamlines is smallest near the particle tips so that inertia leads to larger centrifugal drag forces at the tips and therefore to a larger deformation.
The observed increase of deformation with \Renum is expected to amplify the inward migration of the particles, which may at least partially explain the data in figure \ref{fig:lateral_single} and therefore the inertial focusing in figure \ref{fig:Delta_delta} and \ref{fig:snapshots_density}.

For an even higher capillary number ($\Canum = 0.3$) the deformation is very large, reaching aspect ratios of more than $a / c = 2.5$.
The theory by \citet{barthes-biesel_time-dependent_1981} is not valid anymore.
Interestingly, the \Renum-dependence nearly vanishes then; only in the central region ($z \approx 0$) the deformation increases with \Renum.
The particles are probably already deformed so much that the additional inertial stresses do not contribute noticeably.
It is interesting to note that also particles at the centreplane are deformed, which is caused by their finite size.

The inclination angle of a particle is the angle between its large main axis and the flow direction.
In the small deformation regime the inclination angle assumes a value
\begin{equation}
 \label{eq:inclination_analytical}
 \frac{\theta}{2 \pi} = \frac{\theta_0}{2 \pi} - \frac{15}{16} \Canum
\end{equation}
where $\theta_0 = \pi / 4$ (\emph{i.e.}, $45^\circ$) is the asymptotic value for $\Canum \to 0$ \citep{barthes-biesel_motion_1980}.
The computed inclinations are shown in the right column of figure \ref{fig:deformation+inclination_profiles} for three different capillary numbers.
Since the angles for $z < 0$ are negative, we report only the magnitude and denote it by $\theta$.
Indeed, figure \ref{fig:deformation+inclination_profiles} reveals that all particles at $\lvert z \rvert > r$ have an inclination angle of $\approx 45^\circ$ in the limit of small \Renum and \Canum.
Depending on \Canum, one observes different characteristics of the inclination angle.
For $\Canum = 0.003$, $\theta$ strongly depends on \Renum, especially in the wall vicinity.
However, $\theta$ increases with \Renum and reaches values of about $60^\circ$.
\Citet{laadhari_vesicle_2012} also observed an increase of the inclination angle of steadily tank-treading vesicles with \Renum.
Only in the vicinity of the centreplane ($\lvert z \rvert < r$), $\theta$ is independent of \Renum.
Also for $\Canum = 0.03$ and $0.3$, the angles are strictly increasing with \Renum.
The increase of the local viscosity with \Renum (figure \ref{fig:viscosity_vs_density}) can be explained based on these findings; when the average inclination angle increases with \Renum, the particles assume a larger cross-section in the channel, which in turn leads to a larger flow resistance and therefore dissipation.

As predicted by equation \eqref{eq:inclination_analytical}, the inclination angle also shows a strong $z$-dependence.
Due to the finite size of the particles located near the centreplane of the Poiseuille flow, $\theta$ cannot be directly compared to that in a simple shear flow.
Therefore, the inclination angles in the central region show strong deviations from the analytic solution.

Comparing the data for a single particle and the suspensions, we generally observe that collective effects lead to a decrease of the deformation parameter and an increase of the inclination angle, in particular for larger values of \Renum.
This may have additional non-linear feedback effects on the lateral particle distributions.

\section{Summary and conclusions}
\label{sec_summary}

We have performed three-dimensional computer simulations of a particle suspension using a finite-element-immersed-boundary-lattice-Boltzmann method to investigate the interplay of fluid inertia and particle deformability in a planar Poiseuille flow.
The channel Reynolds number \Renum and the particle capillary number \Canum are used as free control parameters ($\Renum \in [3, 417]$, $\Canum \in [0.003, 0.3]$) while the suspension volume fraction $\phi$ and channel confinement $\chi$ are kept fixed ($\phi = 0.1$, $\chi = 0.20$).
Additionally we have performed reference simulations for a single particle ($\phi = 0.001$) under otherwise identical conditions.

We found that the Segré-Silberberg effect \citep{segre_behaviour_1962, segre_behaviour_1962-1} is suppressed upon an increase of the particle deformability.
While the concentration profile of the nearly rigid particles ($\Canum = 0.003$) shows a central depletion region and a distinct peak in the vicinity of the walls for $\Renum \approx 30$--$200$, the central depletion region for more deformable particles ($\Canum > 0.03$) at fixed \Renum vanishes, and the lateral particle distribution becomes more narrow.
We therefore conclude that the deformability-induced centrewards migration eventually exceeds the inertial lift force.
This view is supported by the behaviour of a single particle; in particular, the particle for $\Renum = 6$ and $\Canum = 0.3$ (\emph{i.e.}, for minimum inertia and maximum deformability effect) assumes a lateral equilibrium position on the centreplane while all other particles end up at finite distances.

Another peculiar suspension behaviour has been found when both \Renum and \Canum are large.
For $\Renum > 200$ and $\Canum > 0.03$, a particle focusing towards the centreplane takes place which is much stronger than the focusing caused by deformability alone (\emph{i.e.} for large \Canum and small \Renum).
This effect is partly caused by the tendency of a single particle to move closer to the centreplane when \Renum becomes larger than about $45$ as already observed by \cite{shin_inertial_2011}.
But also the role of particle-particle interactions at finite volume fractions is important.
We observe a reduction of the lateral particle velocity fluctuations, indicating a decrease of dispersion forces, when both \Renum and \Canum are large.
This is in line with results obtained by \cite{doddi_effect_2008} who described a change of the nature of particle collisions and consequential reduction of dispersion in this parameter regime.
However, to the best of our knowledge, the inertial focusing in a non-dilute deformable particle suspension in Poiseuille flow has not been described before.

We have also seen that the local suspension viscosity decreases with \Canum but increases with \Renum.
The former effect is well-known and caused by reduced flow resistance of soft particles while the latter finding can be explained by analysing the particle properties in the suspension.
Both the particle deformation and inclination increase with \Renum, which is in agreement with results previously published by other groups \citep{doddi_effect_2008, kim_numerical_2012, laadhari_vesicle_2012}.
More strongly inclined particles lead to an apparent growth of the volume fraction, which in turn increases the local dissipation.
The inertia-augmented deformation may be one explanation for the strong inward migration of deformable particles at high Reynolds numbers.

Despite the growth of the local viscosity with \Renum, the inertial focusing of particles towards the centreplane leads to situations where the apparent viscosity of the overall suspension decreases. The decrease of the dissipation due to particles being shifted from the high-dissipation region near the wall to the low-stress region at the centreplane can overcompensate the local viscosity growth when \Renum is increased.

Generally, the interplay of velocity curvature (\emph{i.e.} non-uniform stress), flow confinement, inertia-augmented particle deformation and inertia-dependent particle interactions leads to a complex suspension behaviour and non-homogeneous lateral particle distributions which in turn affects the rheology of the suspension.
It is highly non-trivial to disentangle cause and effect of the observed phenomena in the present geometry.
The large number of parameters forced us to reduce the dimensionality of the parameter space.
Additional studies are required to investigate the relative contributions of the above-mentioned mechanisms and the effect of the confinement and volume fraction on the inertial focusing.
We hope that this work will stimulate further experimental and theoretical investigations of deformability- and inertia-induced effects in suspension flows.

\begin{acknowledgments}
We thank Phillipe Peyla and Michael D.\ Graham for stimulating discussions, NWO/STW (VIDI grant 10787 of J. Harting) for financial support, and SARA Amsterdam for access to high performance computing resources (grant SH-235-13). We also thank the anonymous referees who provided constructive and valuable comments.
\end{acknowledgments}

\end{document}